\documentclass{IEEEtran}

\usepackage{cite}

\ifCLASSINFOpdf
   \usepackage[pdftex]{graphicx}
\else
   \usepackage[dvips]{graphicx}
\fi

\usepackage[cmex10]{amsmath}

\usepackage{xcolor,soul,framed}

\usepackage{array}
\usepackage{multicol}
\usepackage{arydshln}
\usepackage{svg}
\usepackage{cite}
\usepackage{pifont}
\usepackage{multirow}

\usepackage[final]{changes}

\makeatletter
\let\old@ps@headings\ps@headings
\let\old@ps@IEEEtitlepagestyle\ps@IEEEtitlepagestyle
\def\psccfooter#1{%
    \def\ps@headings{%
        \old@ps@headings%
        \def\@oddfoot{\strut\hfill#1\hfill\strut}%
        \def\@evenfoot{\strut\hfill#1\hfill\strut}%
    }%
    \def\ps@IEEEtitlepagestyle{%
        \old@ps@IEEEtitlepagestyle%
        \def\@oddfoot{\strut\hfill#1\hfill\strut}%
        \def\@evenfoot{\strut\hfill#1\hfill\strut}%
    }%
    \ps@headings%
}
\makeatother

\begin{document}

%\section*{Highlights}
%\begin{itemize}
%    \item Metrics identify promising candidates for busbar %splitting based on grid topology and optimal power flow %results
%    \item Grid topology optimization to reduce total generation %costs via busbar splitting
%    \item Convex-quadratic mixed integer grid topology %optimization model with AC-feasibility check of the optimized %topology
%\end{itemize}

%\newpage
%\renewcommand{\baselinestretch}{0.98}


%\onecolumn

\title{Identifying Best Candidates for Busbar Splitting}

%% To specify the authors when (number of affiliations <= 2)
\author{
\IEEEauthorblockN{Giacomo Bastianel, Dirk Van Hertem, Hakan Ergun}
%\IEEEauthorblockN{Giacomo Bastianel, \\ Hakan Ergun}
\IEEEauthorblockA{KU Leuven / Etch - EnergyVille\\
Leuven / Genk, Belgium\\
%\{giacomo.bastianel, hakan.ergun\}@kuleuven.be}
\{giacomo.bastianel, dirk.vanhertem, hakan.ergun\}@kuleuven.be}
\and\\
\IEEEauthorblockN{Line A. Roald\\}
\IEEEauthorblockA{University of Wisconsin - Madison, \\
Madison, WI, USA\\
roald@wisc.edu}
}

\maketitle

\begin{abstract}
Rising electricity demand and the growing integration of renewables are intensifying congestion in transmission grids. Grid topology optimization through busbar splitting (BuS) and optimal transmission switching can alleviate grid congestion and reduce the generation costs in a power system. 
However, BuS optimization requires a large number of binary variables, and analyzing all the substations for potential new topological actions is computationally intractable, particularly in large grids. To tackle this issue, we propose a set of metrics to identify and rank promising candidates for BuS, focusing on finding buses where topology optimization can reduce generation costs. %from a grid topology optimization perspective. 
To assess the effect of BuS on the identified buses, we use a combined mixed-integer convex-quadratic BuS model to compute the optimal topology and test it with the non-linear non-convex AC optimal power flow (OPF) simulation to show its AC feasibility \added{and generation cost reduction compared to the AC-OPF simulations}. By testing and validating the proposed metrics on test cases of different sizes, we show that they are able to identify busbars that reduce the total generation costs when their topology is optimized. Thus, the metrics enable effective selection of busbars for BuS, with no need to test every busbar in the grid, one at a time.   
\end{abstract}


% Use this to place sponsorships
%\thanksto{\noindent Submitted to the 24th Power Systems Computation Conference (PSCC 2026). This work is supported by the Belgian Energy Transition Fund, FOD Economie, project DIRECTIONS, and a Research Foundation – Flanders (FWO) travel grant for Giacomo Bastianel's research visit to UW-Madison.}

%%%% AC stuff

%% SETS

% nodes
\providecommand{\acnodes}{\mathcal{I}}
% branches
\providecommand{\acbranches}{\mathcal{L}}
%switches
\providecommand{\acswitches}{\mathcal{SW}^{ac}}
\providecommand{\switches}{\mathcal{C}}
%topologies
\providecommand{\actopology}{\mathcal{T}^{ac}}
\providecommand{\actopologyrev}{\mathcal{T}^{ac, rev}}
% switch topologies
\providecommand{\acswitchtopology}{\mathcal{T}^{\text{sw,ac}}}
\providecommand{\acswitchtopologyrev}{\mathcal{T}^{\text{sw}^{\text{ac, rev}}}}
\providecommand{\acZILtopology}{\mathcal{T}^{\text{ZIL,ac}}}

%% VARIABLES

\providecommand{\nodevoltage}{V_i}
\providecommand{\acbranchflow}{S_{lij}}

%%%% DC stuff

%% SETS

% nodes
\providecommand{\dcnodes}{\mathcal{E}}
% branches
\providecommand{\dcbranches}{\mathcal{D}}
%switches
\providecommand{\dcswitches}{\mathcal{SW}^{dc}}
%topologies
\providecommand{\dctopology}{\mathcal{T}^{dc}}
\providecommand{\dctopologyrev}{\mathcal{T}^{dc, rev}}
% switch topologies
\providecommand{\dcswitchtopology}{\mathcal{T}^{\text{sw,dc}}}
\providecommand{\dcswitchtopologyrev}{\mathcal{T}^{\text{sw}^{\text{dc}, rev}}}
\providecommand{\dcZILtopology}{\mathcal{T}^{\text{ZIL,dc}}}

% ac nodes built when performing busbar splitting
\providecommand{\acnodesnew}{\mathcal{I'}}
\providecommand{\acZIL}{\mathcal{S}}
\providecommand{\dcnodesnew}{\mathcal{E'}}
\providecommand{\dcZIL}{\mathcal{Q}}

%% VARIABLES

\providecommand{\dcbranchflow}{P_{def}}

%%%% Loads, generators, converters

%% SETS

\providecommand{\acdcconverters}{\mathcal{C}}

\providecommand{\convertertopology}{\mathcal{T}^{\text{cv}}}

\providecommand{\generators}{ \mathcal{G}}

\providecommand{\loads}{\mathcal{M}}

\providecommand{\dcgenerators}{\mathcal{G}^{\text{dc}}}

\providecommand{\dcloads}{\mathcal{M}^{\text{dc}}}

\providecommand{\genconn}{\mathcal{T}^{\text{gen}}}
\providecommand{\dcgenconn}{\mathcal{T}^{\text{gen,dc}}}

\providecommand{\acloadconn}{\mathcal{T}^{\text{load}}}
\providecommand{\dcloadconn}{\mathcal{T}^{\text{load, dc}}}

%% VARIABLES

\providecommand{\genpower}{ S^g_k }
\providecommand{\acloadpower}{ S^m_k }
\providecommand{\dcloadpower}{ P^m_k }
\providecommand{\converteracpower}{ S^c_l }
\providecommand{\converterdcpower}{ P^{c, dc}_l }

% .... continua tu con le variabili... :)

%\setlength{\parindent}{0mm}
\newcounter{model1} \setcounter{model1}{0}
\providecommand{\modelone}[1]{\noindent%
	\refstepcounter{model1}\text{(M1.\arabic{model1})}\\%
}
\providecommand{\modeltwo}[1]{\noindent%
	\refstepcounter{model2}\text{(M2.\arabic{model2})}\\%
}
\providecommand{\modelthree}[1]{\noindent%
	\refstepcounter{model3}\text{(M3.\arabic{model3})}\\%
}

%%%%%%%%%%%%%%%%%%%%%%%
\section{Introduction and motivation}

With the increasing urgency to decrease global CO$_{2}$ emissions, societies are becoming increasingly electrified. As a result, electricity demand is projected to increase significantly \cite{IEA_2023}. %,IEA_2025}.
However, due to the limited transmission capacity and the distributed nature of newly-installed renewable energy sources (RES) capacity~\cite{IRENA}, power systems worldwide are facing increasing congestion~\cite{IEA_2023}. In both Europe~\cite{ACER_congestion} and the US~\cite{Congestion_US_24,Congestion_US_25}, congestion costs have seen a substantial increase in recent years, reaching e.g. 2.774 bn€ in 2024 in Germany~\cite{SMARD}, and 13.5-8.33 bn\$ in 2023 and 2024 for all the US Independent System Operators~\cite{Congestion_US_24,Congestion_US_25}. Currently, power grids are considered an important bottleneck for a carbon-free society, and the deployment of RES requires additional transmission capacity~\cite{IEA_2023}. Significant RES curtailment will take place without effective grid expansion projects \added{aimed to increase the grid transmission capacity}~\cite{JRC_2024}. At the same time, expanding the grid requires worldwide investments in the range of hundreds of billions~€/\$ per year~\cite{IEA_commentary}. Besides economic reasons, grid expansion projects are often delayed due to technical (supply chain issues, lack of workforce, lack of manufacturing capacity, bottlenecks with critical elements, etc.~\cite{IEA_2025_constraints}) and non-technical (public opposition~\cite{NEUKIRCH_Grid_acceptance}) issues. 
Because of all these reasons, system operators are currently aiming to maximize the transmission capacity in the existing grid to reduce existing congestion. \added{For example,} congestion management actions such as power flow control can be performed, \added{among the others,} with phase-shifting transformers and high-voltage direct current links. A broader term to describe these technologies is ``grid-enhancing technologies''~\cite{MIRZAPOUR2024110304}, which also includes flexible AC transmission technologies and dynamic line rating.
Although the attention surrounding these technologies is increasing~\cite{EPRI}, in Europe, they are still primarily applied to maximize cross-border flows, while additional costly remedial measures are employed to address internal grid congestion. 

Consequently, several initiatives in Europe~\cite{nrao} and the US~\cite{EPRI,NewGrid2022} aim to improve the utilization of existing grid infrastructure by actively optimizing and reconfiguring the network topology to mitigate congestion and maximize transfer capacity. However, the available topological actions, such as OTS and substation reconfiguration with BuS are often restricted to well-established, expert-driven measures that system operators have employed for decades, even though the power flows through transmission \replaced{grids are rapidly changing.}{and may not take into account the distributed or variable nature of recently installed RES capacity.}
As a result, potentially effective topological actions remain unidentified, and systematic methodologies to discover them are yet to be established in the scientific literature. 
In particular, while the most common topological actions in industry include substation reconfiguration through BuS, academic papers investigating topology optimization with BuS are relatively new, in part because BuS is computationally highly challenging. Every busbar considered in the topology optimization problem introduces a new degree of freedom and thus has the potential to reduce generation costs compared to a standard OPF \added{by solving congestion, therefore allowing the cheaper generators to be used to their maximum rated power}, but representing multiple possible splitting configurations requires the use of binary variables and thus adds significant computational complexity.
When implementing topology optimization for larger grids, it is therefore necessary to pre-screen busbars to identify promising candidate busbars for BuS.
This paper addresses this issue by proposing metrics to identify efficient topological actions in transmission grids based on the results of optimal power flow (OPF) simulations. Although several metrics (discussed later) have been proposed in the literature, no prior work has systematically assessed their effectiveness in identifying the most suitable busbars for applying topological actions \added{through BuS}. In a rapidly evolving power system, such metrics can support system operators in identifying new impactful topological actions, ultimately maximizing the grid’s power transfer capacity.

% After this introduction and motivation Section, Section~\ref{sec:literature_review} relates this work with the existing literature on the topic, and lists the paper's contributions. Section~\ref{sec:methodology} describes the methodology used in the paper and includes the test cases to refine and test the identified metrics. Finally, Section~\ref{sec:results} discusses the results, and Section~\ref{sec:conclusion} concludes the paper, also highlighting possible future work.

%%%%%%%%%%%%%%%%%%%%%%%
\subsection{Related work and contributions} \label{sec:literature_review}
The concept of topology optimization using optimal transmission switching (OTS) and busbar splitting (BuS) has been utilized for decades to manage contingencies~\cite{Old_OTS}, reduce the total generation costs (OTS~\cite{Fisher2008,Hedman_2009} and BuS~\cite{Heidarifar2014}), relieve grid congestion, and plan maintenance. 
While the first ideas to filter and select the most relevant network elements to be switched were developed decades ago for~\cite{Mazi1986,Makram1989}, and there is a considerable body of literature on OTS~\cite{Hedman_2011}, its extensions~\cite{Hedman_2008} and application related to contingency analysis~\cite{Hedman_2009}, the proposal to use BuS to optimize the substation configuration and achieve reductions in the total generation costs is fairly recent~\cite{Heidarifar2016,Hinneck2021,Heidarifar2021,Morsy2022} and has been applied to hybrid AC/DC grids, too~\cite{Bastianel,MORSY_2025,Bastianel_2025,Van_Deyck}.

Most literature is based on the concept of locational marginal price (LMP), defined as ``\textit{the marginal cost of delivering one more unit of electricity to a specific bus, taking into account transmission losses and network constraints}''~\cite{book_LMP}. LMPs depend on the location where electricity is generated or consumed, and are usually higher in areas that normally import power and lower in
areas that export power~\cite{Kirschen_Strbac}. From a mathematical perspective, LMPs can be derived as the Lagrange multiplier associated with the nodal balance constraint in the OPF formulation~\cite{Kirschen_Strbac}.

References ~\cite{Ruiz_2011} and~\cite{Fuller_2012} propose a heuristic to select the most relevant lines to be switched in the OTS problem based on the difference in LMPs between the two ends of a branch, multiplied by the power flow through the branch. Additional sensitivity criteria for line selection are investigated in~\cite{Ruiz_2012}. Moreover, potential remedial actions are identified by studying congestion zones in~\cite{Khanabadi_2013} for the OTS problem and in~\cite{Heidarifar2016,Zhou2021ACC} for the BuS problem.~\cite{Heidarifar2021} extends the methodology in~\cite{Heidarifar2016} by adding a pre-screening step that uses LMPs associated with active and reactive power and only considers substations with four or more connected lines to identify relevant busbars for OTS. Other references~\cite{Wang2023,Sogol_2021} develop heuristics for selecting the most relevant busbars in BuS based on the load margin~\cite{Wang2023} and for selecting the best breakers in node-breaker modelling~\cite{Sogol_2021} using i) the maximum increase in the congestion rent, ii) the reduction in total generation costs and iii) the largest absolute value of LMP difference across any line in the system. The latter metric is similar to~\cite{Ruiz_2011}, but without considering the power through the lines. 
Finally, ~\cite{Morsy2022} selects buses with ten or more connected components as candidates for topology optimization. %while we selected 
%In previous work~\cite{Bastianel,Bastianel_2025}, we selected the busbars leading to the highest reduction in generation costs in a combined OTS/BuS optimization where only one busbar was allowed to be split at a time. 
\added{The existing above-mentioned metrics to select the most relevant elements for grid topology optimization mainly rely on single $\Delta_{LMP}$s through a branch connected to a given busbar, without including $\Delta_{LMP}$ patterns throughout all the branches connected to it. In addition, except for~\cite{Heidarifar2021}, they rely on the linear DC approximation of the OPF problem, without considering the grid's reactive power and voltage magnitude levels. These factors are becoming increasingly important in modern converter-dominated power systems, and their mismanagement could lead to events such as the Iberian blackout in April 2025~\cite{Iberian_blackout}.}

In previous work~\cite{Bastianel,Bastianel_2025}, we observed that only certain busbars lead to a reduction in generation costs when only one busbar was allowed to be split at a time. However, if applied to a large system, having to exhaustively check all busbars adds a significant computational cost \added{to the optimization problem}. The goal in this paper is therefore to identify a small subset of \replaced{busbars}{bus} that will reduce generation cost if we optimize their topology. \added{These busbars are selected a priori through the proposed metrics, based on the grid topology and on the results of the AC-OPF formulation.}

Table~\ref{table:metrics_literature} summarizes the most relevant metrics emerging from the literature, and whether they were related to OTS or BuS in the aforementioned publications. The new metrics proposed in this work are introduced later in Section~\ref{sec:metrics}.
\begin{table}%[h!]
\caption{Overview of common metrics in the literature related to optimal transmission switching (OTS) and busbar splitting (BuS). ``LMP'' stands for ``locational marginal price''. ``$\Delta_{LMP}$'' stands for the difference in LMPs over a branch.}
\centering
{\fontsize{7pt}{9pt}\selectfont}
\begin{tabular}{l|c|c}
\hline
 Identified metric & OTS-Related & BuS-Related\\
\hline
 ($\Delta_{LMP}$) $\cdot$ (branch power flow)  & \cite{Ruiz_2011,Fuller_2012} & - \\
 Congested \added{lines/}zones identification & \cite{Khanabadi_2013} & \cite{Heidarifar2016,Heidarifar2021,Zhou2021ACC} \\ 
 Load margin & - & \cite{Wang2023} \\ 
 Increase in congestion rent & \cite{Sogol_2021} & - \\ 
 $\Delta_{LMP}$ & - & ~\cite{Heidarifar2021,Sogol_2021} \\
%  Generation costs reduction & ~\cite{Bastianel,Bastianel_2025,weishaoNewAlgorithmRelieving2004} & \ding{51} \\
\# of elements connected to the busbar & - & ~\cite{Heidarifar2021,Morsy2022} \\
 %Performance index of security Margin & ~\cite{weishaoNewAlgorithmRelieving2004,shaoCorrectiveSwitchingAlgorithm2005} & \ding{55} \\
          \hline
\end{tabular}
\label{table:metrics_literature}
\end{table}

\begin{figure*}%[!ht]
    \centering
    \includegraphics[width=0.8\linewidth]{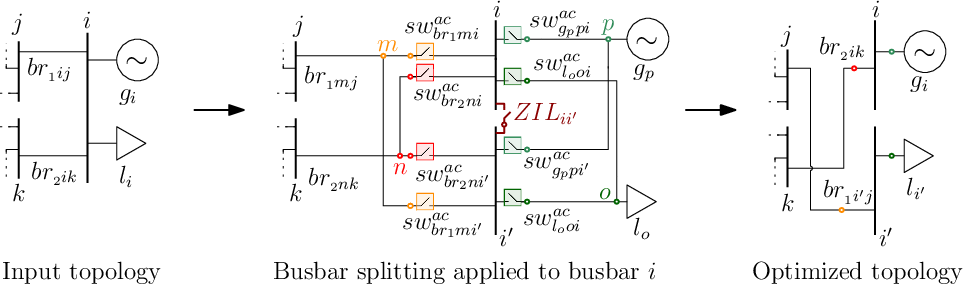}
    \caption{Steps behind the topology optimization model proposed in~\cite{Bastianel,Bastianel_2025}. The selected busbar $i$ (left) is split into two buses $i$ and $i'$, connected through a busbar coupler \added{$ZIL_{ii'}$}. The open/close position of this busbar coupler \added{$ZIL_{ii'}$} is represented  \added{in the optimization model} by a binary variable. Each network element that was connected to busbar $i$ in the input topology is linked to an auxiliary bus (named $m,n,o$ and $p$ in the figure) and can be connected to either one ($i$) or the other part ($i'$) of the split busbar through a switch (center).  \added{Each switch is also represented in the model through a binary variable.} After the optimization, the inactive switches are removed to yield the new topology (right).}
    \label{fig:BuS_model}
\end{figure*}

\subsection{Contributions}
    This paper improves on the state of the art in the field of steady-state grid topology optimization by:
    \begin{itemize}
        \item Identifying and validating metrics to select relevant busbar candidates to perform busbar splitting on AC grids ranging from 39 to thousands of buses, with the final objective of reducing the total generation costs. The pre-emptive selection of relevant candidates for busbar splitting identifies zones where grid topology optimization is relevant without the need to investigate the whole grid. 
        \item Validating the topology optimized with the busbar splitting model introduced in~\cite{Bastianel,Bastianel_2025}, based on a piecewise linear approximation (LPAC) of the AC-optimal power flow (OPF) formulation, with respect to LPAC- and AC-OPF simulations, obtaining feasible and optimal results. 
    \end{itemize}

The remainder of this work is organized as follows.Section~\ref{sec:methodology} describes the existing topology optimization model used in the paper, while Section~\ref{sec:metrics} describes the proposed metrics for selecting the best candidates for \replaced{busbar splitting}{BuS}. Section~\ref{sec:results} discusses the results, withSection~\ref{sec:conclusion} concluding the paper, and highlighting possible avenues for future work.

%%%%%%%%%%%%%%%%%%%%%%%

\section{Preliminaries: Topology Optimization with Busbar Splitting} \label{sec:methodology}
This paper is based on the grid topology optimization model introduced in~\cite{Bastianel,Bastianel_2025}, which includes BuS of selected busbars based on an LPAC-OPF formulation. This model gives rise to a mixed-integer, convex-quadratic optimization problem that can be solved  efficiently compared with a formulation that uses the full non-convex, non-linear AC power flow constraints, while also providing results that are typically AC-feasible~\cite{Bastianel,Bastianel_2025}. 
 
Once we obtain the optimized topology, we evaluate the cost reduction by solving the AC-OPF from ~\cite{PowerModels2018} with the original and optimized topologies. 
For brevity, we refer the reader to ~\cite{PowerModels2018} for a thorough explanation of the detailed \added{LPAC- and} AC-OPF formulations, \added{while here we add the LPAC-BuS model from~\cite{Bastianel}}, and we focus on explaining the logical constraints used to represent the BuS actions in the model to highlight their inherent complexity. 

\added{\subsection{Intuition behind busbar splitting}}
Fig.~\ref{fig:BuS_model} provides an example of a busbar with two branches, a generator and a load.
% \begin{figure*}
%     \centering
%     \includegraphics[width=0.9\linewidth]{Figures/Illustrations/BuS_illustration.eps}
%     \caption{Steps behind the topology optimization model proposed in~\cite{Bastianel,Bastianel_2025}. The selected busbar $i$ (left) is split into two buses $i$ and $i'$, connected through a busbar coupler \added{$ZIL_{ii'}$}. The open/close position of this busbar coupler \added{$ZIL_{ii'}$} is represented  \added{in the optimization model} by a binary variable. Each network element that was connected to busbar $i$ in the input topology is linked to an auxiliary bus (named $m,n,o$ and $p$ in the figure) and can be connected to either one ($i$) or the other part ($i'$) of the split busbar through a switch (center).  \added{Each switch is also represented in the model through a binary variable.} After the optimization, the inactive switches are removed to yield the new topology (right).}
%     \label{fig:BuS_model}
% \end{figure*}
To represent BuS actions for busbar $i$ in Fig.~\ref{fig:BuS_model} (left), we remove all grid elements connected to the busbar and instead connect them to new auxiliary busbars. Then, the original bus is split into two buses $i$ and $i'$. For each component, we introduce a pair of switches ${sw_{\upsilon mi}^{ac}}$ and ${sw_{\kappa mi'}^{ac}}$ to represent which part of the split busbar the component can be connected to. Each component cannot be connected to more than one bus, but can be completely disconnected. We also introduce a binary variable ${sw_{ZILii'}^{ac}}$ to represent the open/close status of the busbar coupler, which connects the busbars $i$ and $i'$.
If the busbar coupler ${sw_{ZILii'}^{ac}}$ is closed, the two parts of the AC busbar are connected, and AC buses $i$ and $i'$ have the same voltage \added{angle and magnitude} values. \added{This reasoning is translated into the optimization model from~\cite{Bastianel,Bastianel_2025} in the following Section.}

 \added{\subsection{Grid topology optimization model }}

 \added{The nomenclature of the network components, topologies and connectivities used throughout the Section are included in Table~\ref{tb:Nomenclature}. Note that by using `reverse' topologies for the branches, power flows in both directions can be represented in the OPF and BuS models~\cite{PowerModels2018}.}

\added{
\begin{table}%[!ht]
	\renewcommand{\arraystretch}{1.1}
	\centering
    \caption{Network components, topologies and connectivities and nomenclature of each network component included in the optimization model. ``BuS'' stands for ``busbar splitting''.} 
    \label{tb:Nomenclature}
    {\fontsize{8pt}{11pt}\selectfont
	\begin{tabular}{m{15 em} l}
		\hline
		Network components &\\
		\hline
            $\acnodes$          & Set of nodes             \\           
            $ \acbranches $     & Set of branches  \\          
            $ \switches $       & Set of busbar couplers  \\          
            $ \generators $     & Set of generators   \\
            %$ \dcgenerators $   & Set of DC generators \\
            $ \loads $          & Set of loads   \\
            $ \acnodesnew $     & Set of auxiliary nodes for BuS \\
            $ \acZIL $          & Set of auxiliary switches for BuS \\
 		\hline
		Topologies and connectivities &\\
		\hline
            $\actopology$         & Branch topologies             \\
            $\actopologyrev$      & Reverse branch topologies     \\ 
            $ \acswitchtopology $ & Switch topologies      \\
            $ \acZILtopology $    & Busbar coupler topologies  \\    
           $ \acloadconn $       & Load connectivity      \\
            $ \genconn $          & Generator connectivity \\
            \hline
		Nomenclature of each network component &\\
            \hline
            $lij~\in~\actopology~\subseteq~\acbranches~\times~\acnodes~\times~\acnodes$ & Branches \\
            $\upsilon ii'~\in~\acZILtopology \subseteq~\switches~\times~\acnodes~\times~\acnodesnew$ & Busbar couplers \\
            $\upsilon mi~\in~\acswitchtopology\subseteq~\acZIL~\times~\acnodes~\times~\acnodesnew$ &  Auxiliary switches for BuS \\
            $lji~\in~\actopologyrev~\subseteq~\acbranches~\times~\acnodes~\times~\acnodes$ & Reverse Branches \\
            $mi~\in~\acloadconn$ & Loads \\
            $gi~\in~\genconn$ & Generators \\
            \hline
	\end{tabular}
    }
\end{table}
}

 \added{The LPAC-BuS model is represented by the equations (M1.1)-(M1.18), as shown in Model 1.
The objective function in (M1.1) consists of i) minimizing the total costs of the $ \generators $ set of generators,
where $P_{k}^{g}$ is the active power injection, and ii) adding a (small) penalty term $c_{ZIL}$ to the busbar couplers $\switches$ to split a busbar only if it actually brings economic benefit. All the generators $k$ in the network are assigned cost parameters $c_{2k}$ (\$/$W^{2}$), $c_{1k}$ (\$/W),  $c_{0k}$ (\$). Note that the quadratic cost term of each generator is assigned to zero.}

 \added{Eq. (M1.2) sets the voltage angle of the reference bus to zero while (M1.3) and (M1.4) limit respectively the voltage magnitude and angles of the other buses between minimum ($\underline{U}^{m}_{i},\underline{\theta}_{i}$) and maximum ($\overline{U}^{m}_{i},\overline{\theta}_{i}$) values. Similarly, the generator setpoints are bounded by $\underline{S}^{g}_{k}$ and $\overline{S}^{g}_{k}$ in (M1.6). The power balance for each node (or substation) is included in (M1.5), with $S^m_l$ being the
nodal demand. The term $Y_{i}^{s} \cdot (1+2\phi_{i})$ refers to the power absorbed by shunt elements connected to the node $i$, which are considered negligible in this paper. $\phi_{i}$ indicates the voltage magnitude change in the LPAC formulation~\cite{LPAC}. The active and reactive power flow through the branches are modeled through (M1.7) and (M1.8), with each branch being represented by a $\pi$-section model~\cite{PowerModels2018}. These active and reactive powers are constrained by (M1.9) and (M1.10), while the piecewise linear approximation of the cosine term $cs$ of $cos(\theta_{n} - \theta_{m})$ for the LPAC formulation~\cite{LPAC} is expressed in (M1.11).}

 \added{Moreover, the reasoning about BuS introduced in the previous Section} is formulated through constraints (M1.12) to (M1.18) and applied to a set of switches $\acswitchtopology$ and busbar couplers $\acZILtopology$, with each switching element represented by a binary variable. (M1.13) and (M1.14) are related to the voltage angles and magnitudes of the two buses at the extremes of each ($sw^{ac}_{\upsilon mi}$) switching element, with $M_{\theta}, M_{m}$ being big-enough M values ($2\pi,1$). (M1.15) and (M1.16) limit the active and reactive powers of all switching elements to its maximum $\overline{P}^{sw,ac}_{\upsilon mi}$, $\overline{Q}^{sw,ac}_{\upsilon mi}$ and minimum values $\underline{P}^{sw,ac}_{\upsilon mi}$, $\underline{Q}^{sw,ac}_{\upsilon mi}$. (M1.7) and (M1.8) are ``exclusivity" constraints. Line switching is allowed to be performed by modeling (M1.7) as an inequality ($\leq$ 1) constraint. As a result, switches $z^{sw,ac}_{\upsilon mi}$ and $z^{sw,ac}_{\kappa mi'}$, connecting each network element to the parts of the split busbar, are allowed to be open simultaneously, therefore removing the network element from the split busbar. Note that (M1.7) imposes that if the busbar coupler is closed, the switch $sw^{ac}_{\kappa mi',t}$ connecting the element to the original busbar is always closed.

\begin{table}%[htbp]
	\renewcommand{\arraystretch}{1.0}
	\centering
	\label{tb:Model1}
        {\fontsize{7.5pt}{12.5pt}\selectfont
	\begin{tabular}{m{28em} l}
		\hline
		 \added{Model 1: Grid topology optimization model, LPAC formulation} &\\
		\hline
		\textbf{Minimize:}\\
		 $\sum_{k \in G} c_{2k}\cdot {\genpower}^{2} + c_{1k}\cdot \genpower + c_{0k} + \sum_{ZILii' \in \acZILtopology} c_{ZIL}\cdot z^{sw,ac}_{ZILii'}$ & \modelone \\
         \\
         \textbf{Subject to}: & \\
	   \textbf{AC bus:}\\		
          $\theta_{r} = 0 $ \label{eq:voltage} & \modelone \\
          $(\underline{U}^{m}_{i} - 1) \leq \phi_{i} \leq (\overline{U}^{m}_{i} - 1)  \quad \forall i \in \acnodes,  $ \label{eq:voltage_magnitudes} & \modelone\\  
          $ \underline{\theta}_{i} \leq \theta_{i} \leq \overline{\theta}_{i}  \quad \forall i \in \acnodes, $ \label{eq:voltage_angles} & \modelone \\
        $\sum_{\substack{k \in \generators_i}} S^g_{k} + \sum_{\substack{l \in \loads_{i}}} S^m_{l} - Y^s_{i}(1+2\phi_{i}) + \sum_{\substack{\upsilon mi \in (\acswitchtopology_{i} \cup \acZILtopology_{i})}} {S^{sw,ac}_{\upsilon mi}} = \sum_{\substack{lij\in \actopology}} S^{ac}_{lij}$  $ \forall i \in \acnodes, $ & \modelone \\   
        \\
          \textbf{Generator}&\\
          $\underline{S}^{g}_{k} \leq S^g_k \leq  \overline{S}^{g}_{k}, \quad \forall k \in \generators \label{eq:gen_limit}$ & \modelone\\
          \\
          \textbf{Branches}&\\
        $P_{lij} = (g_{s,ij} + g_{ij})(1 + 2\cdot\phi_{i}) -g_{ij} (cs_{ij,wt} + \phi_{i} + \phi_{j}) - b_{ij}(\theta_{i}-\theta_{j})  \quad \forall lij\in \actopology \cup \actopologyrev$ \label{eq:Sij}  & \modelone \\
           $Q_{lij} = (b_{s,ij} + b_{ij})(1 + 2\cdot\phi_{i}) -b_{ij} (cs_{ij,wt} + \phi_{i} + \phi_{j}) - g_{ij}(\theta_{i}-\theta_{j})  \quad \forall lij \in \actopology \cup \actopologyrev$ \label{eq:Sij}  & \modelone \\
        $|P_{lij}| \leq  \overline{P}_{lij}  \quad \forall lij \in \actopology \cup \actopologyrev\label{eq:Sijlimit}$ & \modelone\\
        $|Q_{lij}| \leq  \overline{Q}_{lij}  \quad \forall lij \in \actopology \cup \actopologyrev\label{eq:Sijlimit}$ & \modelone\\
        $cs_{ij,wt} \leq 1 - \frac{(1 - cos(\overline{\Delta\theta_{ij}})}{(\overline{\Delta\theta_{ij}})^{2}}(\theta_{i}-\theta_{j})^{2}  \quad \forall lij \in \actopology \cup \actopologyrev$  & \modelone\\
        \\    
        \textbf{Busbar Splitting}&\\
        $ - (1 - z^{sw,ac}_{\upsilon mi}) \cdot M_{\theta} \leq \theta_{m} - \theta_{i} \leq (1 - z^{sw,ac}_{\upsilon mi}) \cdot M_{\theta}, \quad \forall \upsilon mi \in \acswitchtopology \cup \acZILtopology$ \label{diff_leq_M_delta} & \modelone\\
        $ - (1 - z^{sw,ac}_{\upsilon mi}) \cdot M_{m} \leq \phi_{m} - \phi_{i} \leq (1 - z^{sw,ac}_{\upsilon mi}) \cdot M_{m},  \quad \forall \upsilon mi \in \acswitchtopology \cup \acZILtopology $ & \modelone \\
        $z^{sw,ac}_{\upsilon  mi} \cdot \underline{P}^{sw,ac}_{\upsilon  mi} \leq P^{sw,ac}_{\upsilon  mi} \leq z^{sw,ac}_{\upsilon  mi} \cdot \overline{P}^{sw,ac}_{\upsilon mi},  \quad \forall \upsilon mi \in \acswitchtopology \cup \acZILtopology \label{P_ac_sw} $ & \modelone \\
        $z^{sw,ac}_{\upsilon mi} \cdot \underline{Q}^{sw,ac}_{\upsilon mi} \leq Q^{sw,ac}_{\upsilon mi} \leq z^{sw,ac}_{\upsilon mi} \cdot \overline{Q}^{sw,ac}_{\upsilon mi},  \quad \forall \upsilon mi \in \acswitchtopology \cup \acZILtopology \label{Q_ac_sw}$ & \modelone \\
        $({P^{sw,ac}_{\upsilon mi}})^2 + ({Q^{sw,ac}_{\upsilon mi}})^2 \leq z^{sw,ac}_{\upsilon mi}\cdot({\overline{S}^{sw,ac}_{\upsilon mi}})^2,  \quad \forall \upsilon mi \in \acswitchtopology \cup \acZILtopology, \label{S_ac_sw}$ & \modelone \\
        $z^{sw,ac}_{\upsilon mi} + z^{sw,ac}_{\kappa mi'} \leq 1,  \quad \forall (\upsilon mi, \kappa mi') \in \acswitchtopology  \label{z_mn_sw_ots}$ & \modelone  \\
        $z^{sw,ac}_{\kappa mi'} \leq (1 - z^{sw,ac}_{ZILii'}),  \quad \forall (\kappa mi', ZILii') \in \acswitchtopology \cup \acZILtopology \label{z_mn_sw_integer_cut}$ & \modelone  \\
        \hline
	\end{tabular}
    }
\end{table}

\section{Proposed Metrics to Select Candidate Busbars for Busbar Splitting} \label{sec:metrics}
We propose a set of metrics to identify busbars that are likely to lead to cost reductions if their topology is optimized. The goal of the metrics is to avoid an exhaustive search across all buses in the network, as this might lead to prohibitive computational cost. Note that we will still consider cost reductions achieved by splitting a single busbar, as this allows us to test and validate our metrics. We leave BuS optimization with multiple busbars for future work, as this leads to a combinatorial explosion in the possible busbar selections. 

\subsection{Metric design} \label{sec:metric_design}
When designing these metrics, we only use information that can be obtained with comparatively low computational effort, such as from direct analysis of the system topology,e.g., identifying the number of network elements connected to a busbar, or from solving an AC-OPF. We further take inspiration from the metrics previously discussed in Table~\ref{table:metrics_literature} to identify candidate \ \added{lines to be switched in OTS problems and} busbars for splitting in BuS problems. Specifically, we consider the general idea of identifying buses that are associated with congestion, have large differences in LMP to neighboring buses, and where a significant number of elements are connected to the busbar. Based on this, we propose the following three metrics:

\subsubsection{Metric $\phi$ -- Sum of absolute LMP differences} 
This metric is defined as the sum of the absolute differences in LMPs $|\Delta_{LMP}|$ across all branches connected to the selected busbar. This metric rewards buses that have multiple, high $\Delta_{LMP}$ values and is inspired by \textit{congestion zone identification}, $\Delta_{LMP}$, and \textit{\# of elements connected to the selected busbar} metrics.
If a busbar lies within a congestion zone, 
even if one of its branches exhibits a large $\Delta_{LMP}$, the remaining connections are likely to have limited $\Delta_{LMP}$ values, thus limiting the value of BuS to reduce generation costs~\cite{Coffrin_primal_dual}. Moreover, the potential to find topological actions reducing the total costs increases with the number of elements connected to a busbar~\cite{Heidarifar2021,Morsy2022}. Thus, busbars exhibiting large $\Delta_{LMP}$ values over multiple incident branches are promising candidates.

\subsubsection{Metric $\xi$ -- Number of congested branches} This metric evaluates the loading of the branches incident to a busbar and identifies the number of congested branches, inspired by a combination of the \textit{congestion zones identification} and \textit{\# of elements connected to the busbar} metrics. A significant disparity in utilization of branches connected to the same busbar indicates the presence of local bottlenecks, where one or more branches operate closer to their thermal limits while others remain underutilized. Such imbalances suggest opportunities for remedial actions to redistribute flows, alleviate congestion, and therefore reduce generation costs  \added{by trying to use the cheapest generators to their maximum capacity.}

\subsubsection{Metric $\zeta$ -- Binding voltage angle and magnitude limits} 
While a majority of existing work on OTS and BuS metrics has only considered DC-OPF representations, we base our metrics and optimization on AC-OPF (and approximations thereof) that also includes consideration of reactive power and voltage magnitudes. Specifically, the OPF includes voltage magnitude, voltage difference, and angle difference constraints for each branch. 
Branches with binding voltage magnitude or angle difference limits imply congestion and restricted power transfer capability. Identifying busbars connected by branches close to these operational limits highlights locations where topological actions can increase (or not) the grid transmission capacity.

\subsection{Evaluation, testing and validation}

To \emph{evaluate} the metrics, we first solve an AC-OPF with the original topology to determine the generation costs. Then, we solve the LPAC-BuS model with one busbar selected for each run of the topology optimization model, and finally run an AC-OPF with the optimized topology as a feasibility check and to evaluate the cost reduction.

We use a \emph{testing phase} where we leverage results from two small test systems to assess what combination of metrics effectively identifies the most relevant candidates for BuS. We first evaluate the results of enabling BuS for each busbar by computing both the metrics and the corresponding cost reduction for that busbar. Then, we leverage the results to find effective combinations of metrics.  Since it is possible that we overfit our metrics to the specific test cases during the testing phase, we then implement a \emph{validation phase}. In the validation phase, we apply the combination of metrics to two larger test cases and analyze their performance on this previously unseen data. 

Table~\ref{table:test_cases} lists the test cases from pglib~\cite{pglib} used in the \textit{testing} and \textit{validation} steps, including the number of generators, buses, branches, and loads in each test case.  \added{They range from 39 to 3374 buses.} 

\begin{table}%[htbp]
\caption{Test cases from pglib~\cite{pglib} used in this work to test and validate the proposed metrics.}
\fontsize{7pt}{9pt}\selectfont
\centering
\begin{tabular}{cc|cccc}
\hline
Test case & Name in pglib~\cite{pglib} & Generators & Buses & Branches & Loads \\ 

\hline                 
39-bus   & case39\_epri  & 10 & 39  & 46  & 21 \\
118-bus  & case118\_ieee & 54 & 118 & 186  & 99 \\
793-bus  & case793\_goc  & 214 & 793  & 913  & 507 \\
3374-bus & case3375wp\_k & 596 & 3374 & 4161 & 2434 \\
\hline
\end{tabular}
\label{table:test_cases}
\end{table}

 \added{\subsection{Simulations setup} \label{sec:test_cases}}
The results of the OPF and LPAC-BS models are computed using a MacBook Pro with chip M1 Max and 32 GB of memory, using Gurobi v12.0, MIP gap 0.01\%, maximum simulation time  for LPAC-BS, and Ipopt with linear solver ma97 for the OPF model.
%%%%%%%%%%%%%%%%%%%%%%%
\section{Results and discussion} \label{sec:results}

\subsection{Testing phase results}
We perform the testing phase using the
39-bus and 188-bus test cases introduced in Table~\ref{table:test_cases}.
%Using the 39-bus and 188-bus test case introduced in table~\ref{table:test_cases}, we apply the LPAC-BuS model to each busbar separately without candidate pre-filtering, and analyze the effectiveness of the proposed metrics in identifying the most relevant candidates for BuS. After running the LPAC-BuS model, we fix the optimized topology and run feasibility checks (FC) with the AC- and LPAC-OPF formulations.
%\footnote{The LPAC-OPF model is obtained by removing the BS-related constraints and is formulated by Coffrin and Van Hentenryck~\cite{LPAC}.}. 
%These AC-FC and LPAC-FC are useful to test the results of the LPAC-BuS model for additional test cases compared to the ones used in~\cite{Bastianel,Bastianel_2025}.

\subsubsection{Accuracy and Computational Efficiency} 
Using the 39-bus test case as an example, we first make some basic observations about the accuracy and computational effort associated with our approach. 
Fig.~\ref{fig:case_39_one_bus_per_time} plots the $\phi$-metric of each busbar (representing the sum of absolute LMP differences across all the branches connected to the selected busbar) against the reduction in generation cost achieved by splitting that bus. Each busbar is represented by two dots, where the blue dot represents the results achieved with an LPAC-OPF feasibility check (FC) and the green dot represents the results achieved with the AC-OPF FC. The FCs are run on the optimized topology computed through the LPAC-BuS model.

We observe that there are some differences between the two models, but that the trends (e.g. busbars with the highest $\phi$ values and the highest generation cost reductions) are similar. This shows that topology optimization through the LPAC-BuS  \added{model provides results that are sufficiently accurate to reduce the total costs of an AC-FC compared to the AC-OPF of the original topology. Therefore, the LPAC-BuS model computes topologies that are optimal and lead to a reduction in total costs compared to the original topology, even for AC-OPF formulations.}

Regarding computational efficiency, the whole optimization process (splitting one busbar per time with LPAC-BuS, plus AC-FC and LPAC-FC) took 11.4~s, confirming the efficiency of the proposed LPAC-BuS model.

\subsubsection{Testing of Metrics on 39-bus test case}
\begin{figure}%[h!]
        \centering
        \includegraphics[width=0.95\linewidth]{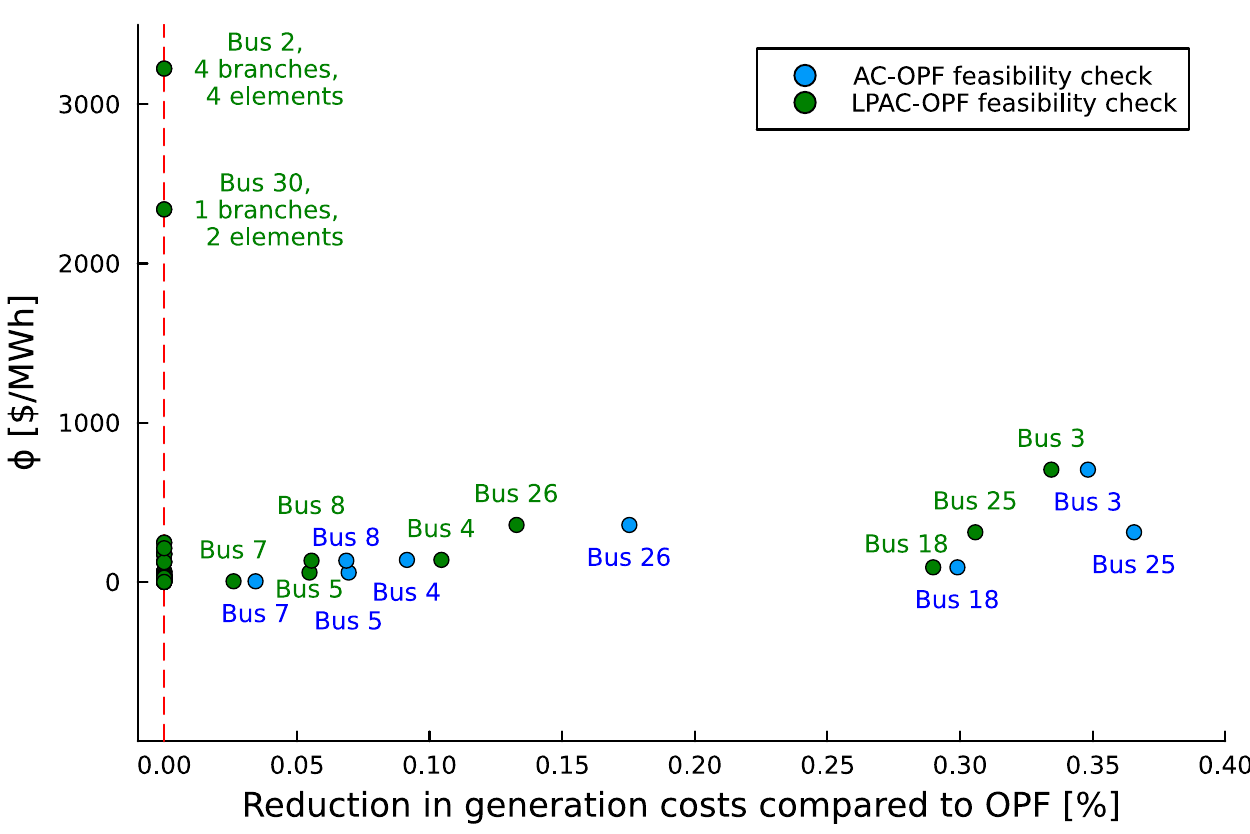}
        \caption{For each bus in \textit{39-bus test case}, the optimal topology from the LPAC-BuS model is tested for an ``AC''-OPF and ``LPAC''-OPF feasibility check (FC). The differences in total costs of the FC compared to the original OPF are plotted on the x-axis, while the y-axis orders the busbar according to each bus' sum of the absolute differences in LMPs over the branches connected to it ($\phi$). The LPAC-FC and AC-FC see reductions in costs for the same buses.}
        \label{fig:case_39_one_bus_per_time}
    \end{figure}

From Fig.~\ref{fig:case_39_one_bus_per_time}, we observe that BuS of buses \textit{25, 3, 18, 26} and \textit{4} lead to the biggest reductions in generation costs, while, interestingly, the two buses with highest $\phi$-values, buses \textit{30} and \textit{2} do not reduce total costs.

Further analysis of the results yields some interesting insights. 
First, the branch connecting busbars \textit{2} and \textit{30} has the highest LMP difference $\Delta_{LMP}$. However, when either of these buses is allowed to be split, the optimal topology is the same as the original one. Thus, there is no change in generation cost. This finding confirms that the widely used metric $\Delta_{LMP}$ is insufficient, and that we need more (or combined) metrics to select relevant busbars for topology optimization. Further, busbar \textit{30} has only 2 elements connected to it, and its voltage magnitude $U^{m}_{30}$ is at the upper limit 1.1. This suggests that it is important to also consider the \textit{\# of elements connected to the selected busbar} and the $\zeta$-metric representing the presence of \textit{binding voltage angle and magnitude limits}.
Busbars \textit{2} has 2 congested lines (used almost at full capacity) connected to it, confirming that the $\xi$-metric representing the \textit{number of congested lines} might be important,  \added{but not when the busbar resides in a congested zone, as we commented in Section~\ref{sec:metric_design}.}

\subsubsection{118-bus test case} 
We apply the previous findings to select busbars for splitting in the 118-bus case. Table~\ref{table:118_results} shows the value of the different metrics as well as the cost reduction achieved with BuS for the buses with the highest $\phi$-metric values (top) and busbars that lead to a cost decrease of $>$0.05\%, but are not in the top-10 $\phi$-ranked busbars (bottom). 

\begin{figure*}
    \centering
    \includegraphics[width=0.85\linewidth]{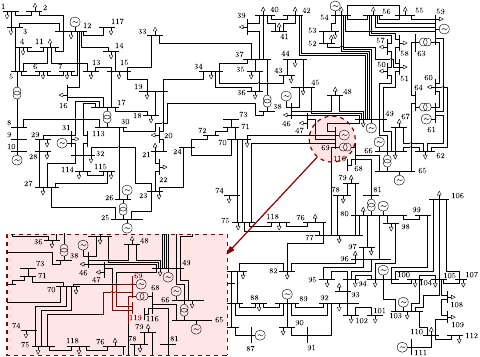}
    \caption{Topology optimization model applied to busbar 69 in the \textit{118-bus test case}. In the optimized topology, busbars \textit{47} and \textit{49} are decoupled from busbar \textit{69} and the neighboring busbars, leading to a reduction in the total generation costs of 0.380\% compared to the original AC-OPF.}
    \label{fig:69_on_118}
\end{figure*}

We observe that the busbars with high $\phi$, many \textit{connected branches/elements} and without \textit{binding voltage limits} ($\zeta$) lead to a cost decrease compared to the AC-OPF. Importantly, 4 out of 5 busbars with a cost reduction larger than 0.1\% are included among the busbars with the highest $\phi$ values.  \added{The highest cost decrease is for busbar \textit{69}, whose optimized topology is shown in Fig.~\ref{fig:69_on_118}. In this optimized topology, power can directly flow from busbar~\textit{49} to~\textit{47}, as the branches connecting them are decoupled from the original busbar~\textit{69}.}

 Among the buses in the bottom part of Table~\ref{table:118_results}, we observe that busbars \textit{80} and \textit{42} are connected to several branches. The other busbars decrease cost by not reconnecting branches to the split busbar, essentially performing transmission line switching rather than BuS. They are therefore more complex to identify with our metrics. 
Finally, all the busbars with a cost decrease of $>$0.05\% are within the first half of the 118 $\phi$-ranked busbars, and two of them have $\phi$ above its average value of 219.13 [\$/MWh].

Note that for this test case, the entire optimization process took 214.7~s. Based on these results, we consider \textit{busbars with several connected branches/elements} and a \textit{$\phi$ above average}  \added{as metrics to select relevant busbars} in the \textit{validation} step. 

\begin{table}%[htbp]
\caption{Metrics of the 10 busbars in the 118-bus test case with the highest sum of the absolute differences in LMPs over their branches ($\phi$), and of the busbars with the highest cost decrease wrt AC-OPF, not in the top 10 $\phi$-ranked busbars. Busbars at voltage limit ($\zeta$) do not have a cost decrease, while there are limited congested ($>$ 85\% utilization) branches ($\xi$).}
\fontsize{7pt}{9pt}\selectfont
\centering
\begin{tabular}{ccccccc|c}
\hline
Busbar& $\phi$  & $\phi$ & \# of & $\xi$&\# of & $\zeta$ & \% Cost \\
ID & [\$/MWh] & rank & branches& [-] & elements & [-] & decrease \\

\hline                 
69  & 2253.98 & 1 & 6 & 2 & 7 & - & 0.380 \\
49  & 1836.24 & 2 & 12& 1 & 14& - & 0.344 \\
100 & 1433.50 & 3 & 8 & 2 & 10& $U^{m}$,1.1 & 0.000 \\
59  & 1168.27 & 4 & 7 & 0 & 9 & - & 0.156 \\
47  & 827.49  & 5 & 3 & 1 & 4 & - & 0.000 \\
66  & 665.04  & 6 & 5 & 0 & 7 & $U^{m}$,1.1 & 0.000 \\
103 & 570.94  & 7 & 4 & 1 & 6 & - & 0.039 \\
56  & 559.88  & 8 & 6 & 0 & 8 & - & 0.017 \\
65  & 552.91  & 9 & 4 & 0 & 5 & - & 0.000 \\
77  & 509.41  & 10 & 7 & 0 & 9 & - & 0.132\\
\hdashline 
80 & 371.05 & 19 & 8 & 0 & 10& - & 0.143 \\
42 & 326.99 & 22 & 4 & 0 & 6 & - & 0.098 \\
44 & 115.77 & 57 & 2 & 0 & 3 & - & 0.097 \\
45 & 192.69 & 41 & 3 & 0 & 4 & - & 0.091 \\
43 & 206.29 & 37 & 2 & 0 & 3 & - & 0.064 \\

\hline
\end{tabular}
\label{table:118_results}
\end{table}

% \begin{figure}[h!]
%        \centering
%        \includesvg[width=0.97\linewidth]%{Figures/case_118/case_118.svg}
%        \caption{IEEE 118 test case}
%\end{figure}

\subsection{Validation phase results}
In this step, we apply the metrics discussed above to identify busbars for splitting in the 793-bus and 3374-bus cases listed in Table~\ref{table:test_cases}. 

\subsubsection{793-bus test case}
We first select the 15 busbars with the highest $\phi$-values and remove those with fewer than 4 elements connected to them.
As a result, the 8 busbars on the top part of Table~\ref{table:793_identified_metrics} are selected. Second, we select the 7 busbars with the highest \textit{\# of elements and branches connected to them}, and have a $\phi$ higher than the mean $\phi$ value. These busbars are displayed on the bottom part of Table~\ref{table:793_identified_metrics}. We ensure that all the selected busbars are connected to at most one congested line ($\xi$-metric $\leq 1$) and have no binding voltage angle or magnitudes constraints ($\zeta$-metric $= 0$). 

\begin{table}%[h!]
\caption{Identification of 15 candidates for busbar splitting in the 793-bus test case with the proposed metrics, and related cost decrease. 12 optimized topologies out of 15 lead to a cost decrease in the AC-OPF.}
\fontsize{7.5pt}{9pt}\selectfont
\centering
\begin{tabular}{cccccc|c}
\hline
Busbar& $\phi$ & $\phi$ & \# of & $\xi$&\# of & \% Cost \\
ID & [\$/MWh] & rank & branches& [-] & elements & decrease \\
\hline                 
693	& 2602.76 & 1	& 5 & 1 & 6 & 0.376 \\
470	& 1630.23 & 2	& 3 & 1 & 4 & 0.650 \\
406	& 1447.60 & 6	& 4 & 1 & 4 & 0.000 \\
649	& 1230.62 & 9	& 4 & 0 & 5 & 0.274 \\
677	& 1173.01 & 11  & 3	& 1 & 4 & \replaced{0.000}{infeasible} \\
171	& 970.89  & 13	& 3 & 1 & 4 & 0.141 \\
190	& 919.09  & 15	& 5 & 1 & 6 & 0.000 \\
23	& 802.69  & 18	& 5 & 1 & 6 & 0.972 \\
\hdashline 
587	& 304.01  & 48	& 6 & 0 & 7 & 1.056 \\
116	& 196.95  & 68	& 5 & 0 & 6 & 0.559 \\
73	& 494.92  & 30	& 6 & 0 & 6 & 0.024 \\
373	& 378.82  & 40	& 4 & 0 & 6 & 0.151 \\
655	& 467.55  & 31	& 6 & 0 & 6 & 0.319 \\
612	& 105.87  & 116 & 6 & 0 & 6 & 0.041 \\
561	& 125.52  & 101 & 5 & 0 & 5 & 0.363 \\
\hline
\end{tabular}
\label{table:793_identified_metrics}
\end{table}

The rightmost column of Table~\ref{table:793_identified_metrics} shows the cost decrease for the identified busbars. As for the 118-bus case, selecting busbars with high $\phi$ values and larger \textit{\# of connected branches/elements} leads to cost decreases for several busbars, namely \textit{693, 470, 649, 171} and \textit{23}. At the same time, the 7 busbars selected using the \textit{\# of branches/elements} and $\phi$ values above the mean-$\phi$ lead to cost reductions as well. 

\begin{figure}%[h!]
        \centering
        \includegraphics[width=0.95\linewidth]{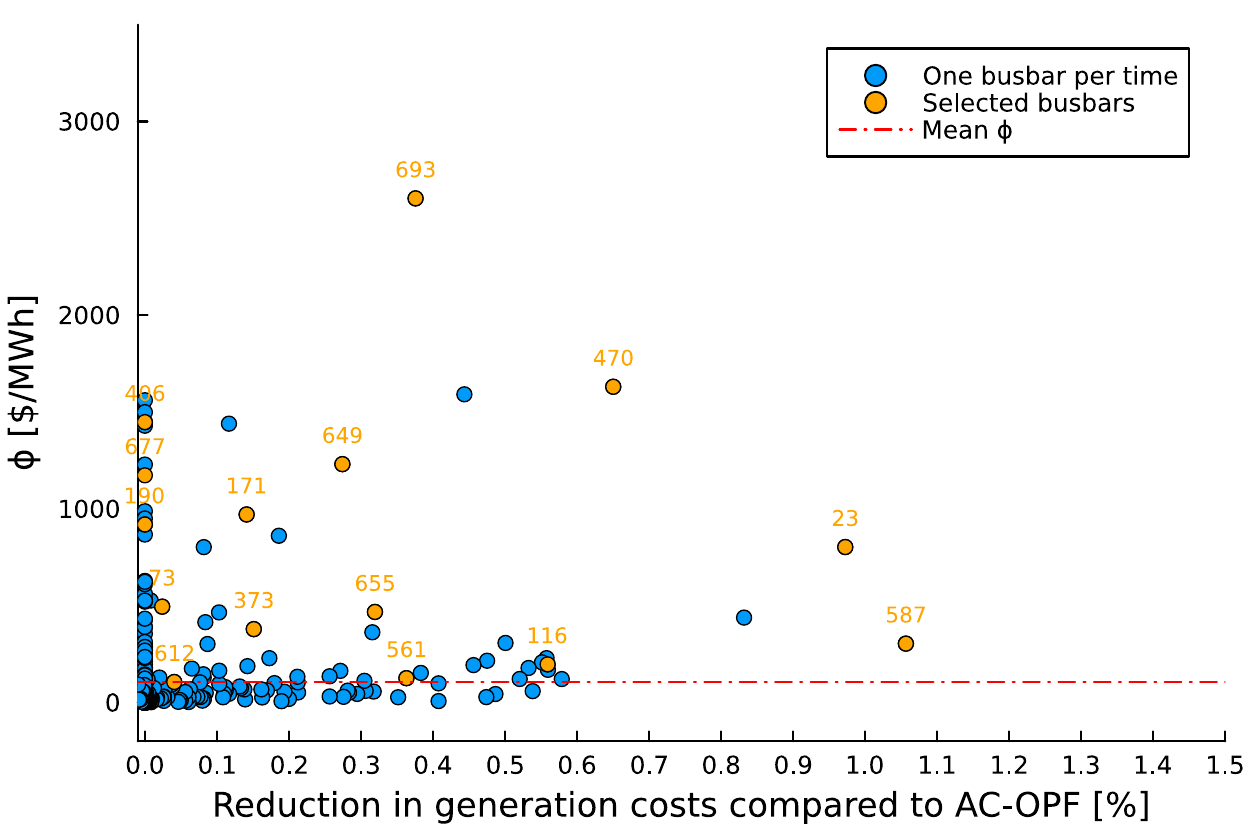}
        \caption{Cost decrease of the AC-OPF feasibility check for the optimized topology of the selected busbars compared to the results obtained by splitting one busbar at a time for the \textit{793-bus test case}. Most of the busbars selected with the identified metrics do lead to a reduction in generation costs. The busbars with the highest cost decreases tend to have a sum of the absolute differences in LMPs across the branches connected to them ($\phi$) higher than the mean $\phi$ value.}
        \label{fig:case_793_one_bus_per_time}
    \end{figure}

To validate whether we identified the ``best'' busbars for splitting, we validate our results by splitting every busbarin the system and checking the cost reduction. 
Fig.~\ref{fig:case_793_one_bus_per_time} shows the results obtained by splitting one busbar at a time for all busbars in the 793-bus system. We observe that the proposed metrics identify 3 of the 4 busbars that lead to the largest reductions in generation costs, as well as several other buses with high cost reductions. We note that busbar \textit{444}, which yields the third-best cost reduction if split, has only two connected branches, both of which are utilized above 80\% of their rated capacity (one at 100\%). As with most busbars in the bottom part of the previous Table~\ref{table:118_results}, not reconnecting a branch to the split busbar is leading to \textit{444}'s reduction in costs, suggesting that line switching rather than BuS is what achieves the cost reduction.

For this test case, splitting one busbar per time took 68102~s (19.14~h), 67639~s for the LPAC-BuS and 463.3~s for the AC-FC, while the total computational time to optimize only the the selected buses is considerably lower, with 3523.3~s (0.98~h).

\subsubsection{3374-bus test case}
Finally, we use the same metrics as in the previous case to identify relevant candidates for BuS for the 3374-bus test case, too. %As this test case is considerably larger than the 793-bus, we consider 20 busbars. 
On the top part of Table~\ref{table:3374_identified_metrics}, the first 13 busbars are selected by considering the 20 busbars with the highest $\phi$-values, and including the only ones with the highest $\phi$ and \textit{\# of connected elements/branches}. Note that busbars \textit{679, 611} and \textit{665} are selected even if they have a low number of branches, as their $\phi$-values are high. In particular, busbars \textit{679} and \textit{611} have $\xi = 1$ and only two branches, as busbar \textit{444} in the previous case, possibly leading to potential for line switching. 
The bottom part of Table~\ref{table:3374_identified_metrics} shows 7 busbars selected based on their \textit{\# of connected elements/branches} and their above-average $\phi$ values. The mean $\phi$ value in this case is 483.55 [\$/MWh].

\begin{table}%[h!]
\caption{Identification of 20 candidates for busbar splitting in the 3374-bus test case with the proposed metrics, and related cost decrease.}
\fontsize{7.5pt}{9pt}\selectfont
\centering
\begin{tabular}{cccccc|c}
\hline
Busbar& $\phi$ & $\phi$ & \# of & $\xi$&\# of & \% Cost \\
ID & [\$/MWh] & rank & branches& [-] & elements & decrease \\

\hline                 
670	   &  71334.40 &	1   &   5	& 1 &	10&  0.048 \\
671	   &  57530.95 &	2   &   5	& 1 &	11&  0.004 \\
679	   &  47530.03 &	3   &   2	& 1 &	3 &  0.000 \\
611	   &  36048.14 &	4   &   2	& 1 &	3 &  0.000 \\
665	   &  30099.63 &	5   &   3	& 0 &	4 &  0.079 \\
254	   &  26184.71 &	6   &   4	& 1 &	9 &  0.067 \\
261	   &  21423.57 &	7   &   5	& 1 &	6 &  0.000 \\
441	   &  18150.84 &	8   &   6	& 0 &	6 &  0.001 \\
460	   &  14854.40 &	11  &	4	& 0 &	4 &  0.072 \\
499	   &  14425.17 &	12  &	5	& 0 &	6 &  0.000 \\
439	   &  14047.40 &	13  &	7	& 0 &	8 &  0.063 \\
498	   &  11704.58 &	16  &	5	& 0 &	6 &  0.056 \\
440	   &  10713.17 &	18  &	5	& 0 &	5 &  0.000 \\
\hdashline
10147	& 827.51 &	355 &	13	& 0 &	13    &  0.000 \\
10153	& 901.66 &	320 &	13	& 0 &	13    &  0.000 \\
10079	& 5648.17 &	43  &	12	& 2 &	10    &  0.000 \\
10145	& 569.98 &	521 &	12	& 0 &	12    &  0.000 \\
10063	& 486.84 &	602 &	12	& 0 &	12    &  0.000 \\
10081	& 1463.67 &	191 &	12	& 0 &	11    &  0.000 \\
10158	& 543.18 &	544 &	12	& 0 &	12    &  0.000 \\
\hline
\end{tabular}
\label{table:3374_identified_metrics}
\end{table}

\begin{figure}%[htbp]
        \centering
    \includegraphics[width=0.9\linewidth]{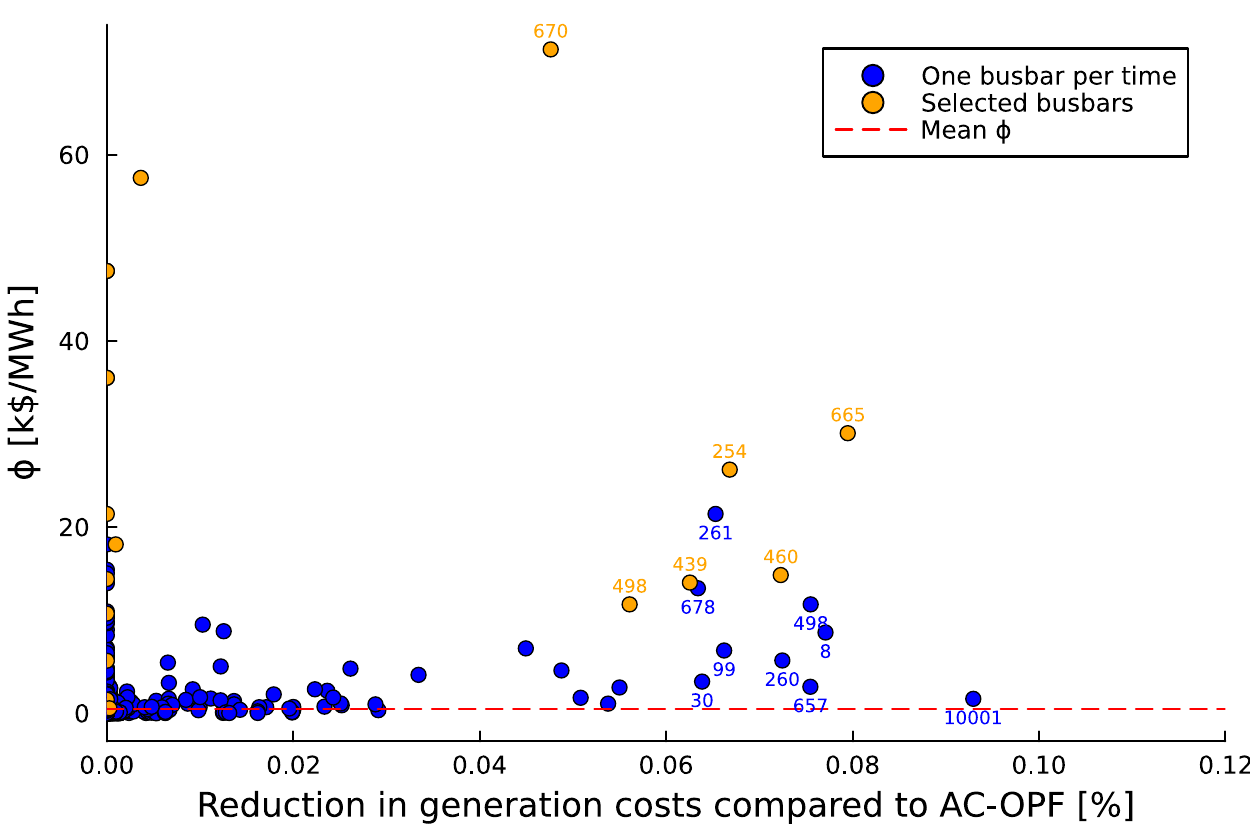}
        \caption{Cost decrease of the AC-OPF feasibility check for the optimized topology of the selected busbars compared to the results obtained by splitting one busbar at a time for the \textit{3374-bus test case}. The proposed metrics identify several busbars leading to the highest reductions in generation costs.}
    \label{fig:case_3374_one_bus_per_time}
    \end{figure}

Table~\ref{table:3374_identified_metrics} and Fig.~\ref{fig:case_3374_one_bus_per_time} show that 8 of the selected busbar leads result in reductions in generation costs compared to the AC-OPF. Also in this case, part of the selected busbars, i.e., \textit{498}, \textit{665}, \textit{439}, and \textit{670}, are among those that lead to the largest cost decrease. Unlike case 793-bus, the busbars in the bottom part of Table~\ref{table:3374_identified_metrics} do not lead to any cost reduction, even though they have several lines and elements connected to them. This result can be attributed to the lower cost reductions relative to previous test cases, indicating that large cost reductions are not achievable in this specific test case, and that identifying relevant busbars for BuS is therefore more challenging. For example, busbar \textit{10001} yields the largest cost reduction, even though it has 4 non-congested branches connected to it and a low $\phi$, making it difficult to identify as a good candidate from the OPF results.  \added{In addition, out of the busbars with high $\phi$-values and low number of branches (\textit{679, 611} and \textit{665}), only \textit{665} leads to a cost decrease if split.} \added{We leave further refinements of the metrics to efficiently select busbars with similar characteristics for future work.}
Nevertheless, Fig.~\ref{fig:case_3374_one_bus_per_time} still confirms the validity of the proposed metrics in identifying relevant busbars for BuS,  \added{with several selected busbars being among the ones leading to the highest cost reductions compared to the AC-OPF simulation}.

Finally, optimizing the selected buses took 13582.7~s (3.77~h), or an average of 11.32~min per bus. Therefore, the proposed metrics select relevant busbars substantially faster than splitting one busbar at a time, which took 374.98~h (15.62~days).

%%%%%%%%%%%%%%%%%%%%%%%
\section{Conclusion and future work} \label{sec:conclusion}
This paper identifies relevant metrics to determine the best busbar candidates for busbar splitting based on the results of optimal power flow simulations. We propose new metrics and combine them with existing ones in the literature.
Through a \textit{testing} step to refine the metrics and a \textit{validation} step, we show that the proposed metrics effectively identify busbars whose optimized topology leads to lower generation costs compared to an AC-OPF simulation.

Specifically, the results show that our newly proposed $\phi$-metric, defined as the \textit{sum of the absolute differences in LMPs over the branches connected to a busbar} ($\phi$), in combination with the \textit{\# of elements connected to the busbar}, can help identify the busbars that achieve the highest cost reductions. Further, the presence of \textit{binding voltage angle and magnitude limits} ($\zeta$-metric) indicates that busbar splitting will likely \emph{not} reduce cost. Some of the buses that were not identified by our metrics achieved cost reductions through transmission line switching rather than busbar splitting.

Given the flexibility of the proposed grid topology optimization model in including more than one busbar, future work will include testing combinations of busbars to achieve further reductions in generation costs. 
Moreover, the proposed metrics can be used in grid planning side to investigate the possibility of retrofitting existing substations to effective topological actions when congestion takes place in the grid, or new substations or RES capacity are installed in the grid.

\section*{Acknowledgement}
This work is supported by the Belgian Energy Transition Fund, FOD Economie, project DIRECTIONS, and a Research Foundation – Flanders (FWO) travel grant for Giacomo Bastianel's research visit to UW-Madison.

\bibliography{Bibliography}
\bibliographystyle{ieeetr}

\end{document}